\documentclass{article}

\begin{document}

\begin{flushright}
CAMS/00-06\\
\end{flushright}
\vspace{1cm}\vspace{1cm} \baselineskip=16pt

\begin{center}
\baselineskip=16pt

\centerline{{\Large \textbf{Deforming Einstein's Gravity}}}

\vskip1cm\centerline{Ali H. Chamseddine}

\centerline\emph{Center for Advanced Mathematical Sciences (CAMS)}

\centerline\emph{and}

\centerline\emph{Physics Department}

\centerline\emph{\ American University of Beirut, Lebanon}
\end{center}

\vskip1 cm \centerline{\bf ABSTRACT} A deformation of Einstein Gravity is
constructed based on gauging the noncommutative $ISO(3,1)$ group using the
Seiberg-Witten map. The transformation of the star product under
diffeomorphism is given, and the action is determined to second order in the
deformation parameter. \vfill\eject

\section{Introduction}

Open string theories as well as D-branes in the presence of background
antisymmetric $B$-field give rise to noncommutative effective field theories
\cite{CDS}-\cite{SW}. This is equivalent to field theories deformed with the
star product \cite{H},\cite{FFZ}. The primary example of this is
noncommutative $U(N)$ Yang-Mills theory \cite{CDS},\cite{SW}.

In recent works \cite{sheikh},\cite{wess}, it was argued that the
gravitational field gets deformed and becomes complex \cite{chams}. The
hermitian metric \cite{Einstein}, includes both the symmetric metric and an
antisymmetric tensor. The analysis of the linearized theory showed that the
theory is consistent, however, further work is needed to show that this is
maintained at the non-linear level, a basic problem faced by all theories of
nonsymmetric gravity \cite{Deser}-\cite{clayton}. For this to happen, it is
essential that the diffeomorphism invariance of the real theory is
generalized to the complex case. This can happen when both the
diffeomorphism transformations and the abelian gauge transformations of the
antisymmetric tensor combine to form complex diffeomorphism. The need for
this is that gauge symmetry prevents the ghost degrees of freedom present in
the antisymmetric tensor from propagating.

The main argument for considering complex vielbein and gauged $U(1,D-1)$ is
that for noncommutative Yang-Mills theory it is only possible to gauge the $%
U(N)$ Lie algebras \cite{SW}. Reality conditions necessary to consider $SO(N)
$ or $SP(N)$ Lie algebras are not possible. This obstacle was overcome \cite
{sheikh},\cite{wess}, by realizing that it is possible to define subgroups
of orthogonal and symplectic subalgebras of noncommutative unitary gauge
transformations even though the gauge fields are not valued in the
subalgebras of the $U(N)$ Lie algebra. What makes this possible is that one
can generalize the reality condition to act on the deformation parameter.
Thus the gauge fields are taken to be functions of the deformation
parameters $\theta $ and the expansion in terms of the non deformed fields
is given by the Seiberg-Witten map. To construct a noncommutative
gravitational action in four dimensions one proceeds as follows. First the
gauge field strength of the noncommutative gauge group $SO(4,1)$ is taken.
This is followed by an Inon\"{u}-Wigner contraction to the group $ISO(3,1)$,
thus determining the dependence of the deformed vierbein on the undeformed
one. At this stage the construction of the deformed curvature scalar becomes
straightforward. The deformed action is computed up to second order in $%
\theta $. The plan of this paper is as follows. In section two we review the
conditions allowing  to deal with noncommutative $SO(N)$ algebras, derive
the noncommutative gauge field strengths, perform the group contraction and
give the deformed curvature scalar. In section three we expand the action to
second order in $\theta $. Section four is the conclusion.

\section{Noncommutative Gauging of SO(4,1)}

The $U(N)$ gauge fields are subject to the condition $\widehat{A}_{\mu
}^{\dagger }=-\widehat{A}_{\mu }.$ Such condition can be maintained under
the gauge transformations
\[
\widehat{A}_{\mu }^{g}=\widehat{g}*\widehat{A}_{\mu }*\widehat{g}_{*}^{-1}-%
\widehat{g}*\partial _{\mu }\widehat{g}_{*}^{-1}
\]
where $\widehat{g}*\widehat{g}_{*}^{-1}=1=\widehat{g}_{*}^{-1}*\widehat{g}$
. Therefore, we first introduce the gauge fields $\widehat{\omega }_{\mu
}^{AB}$, subject to the conditions \cite{sheikh},\cite{wess}
\begin{eqnarray*}
\widehat{\omega }_{\mu }^{AB\dagger }\left( x,\theta \right) &=&-\widehat{%
\omega }_{\mu }^{BA}\left( x,\theta \right) \\
\widehat{\omega }_{\mu }^{AB}\left( x,\theta \right) ^{r} &\equiv &\widehat{%
\omega }_{\mu }^{AB}\left( x,-\theta \right) =-\widehat{\omega }_{\mu
}^{BA}\left( x,\theta \right)
\end{eqnarray*}
Expanding the gauge fields in powers of $\theta $, we have
\[
\widehat{\omega }_{\mu }^{AB}\left( x,\theta \right) =\omega _{\mu
}^{AB}-i\theta ^{\nu \rho }\omega _{\mu \nu \rho }^{AB}+\cdots
\]
The above conditions then imply the following
\[
\omega _{\mu }^{AB}=-\omega _{\mu }^{BA},\quad \omega _{\mu \nu \rho
}^{AB}=\omega _{\mu \nu \rho }^{BA}
\]
A basic assumption to be made is that there are no new degrees of freedom
introduced by the new fields, and that they are related to the undeformed
fields by the Seiberg-Witten map \cite{SW}. This is defined by the property
\[
\widehat{\omega }_{\mu }^{AB}\left( \omega \right) +\delta _{\widehat{%
\lambda }}\widehat{\omega }_{\mu }^{AB}\left( \omega \right) =\widehat{%
\omega }_{\mu }^{AB}\left( \omega +\delta _{\lambda }\omega \right)
\]
where $\widehat{g}=e^{\widehat{\lambda }}$ and the infinitesimal
transformation of $\omega _{\mu }^{AB}$ is given by
\[
\delta _{\lambda }\omega _{\mu }^{AB}=\partial _{\mu }\lambda ^{AB}+\omega
_{\mu }^{AC}\lambda ^{CB}-\lambda ^{AC}\omega _{\mu }^{CB}
\]
and for the deformed field it is
\[
\delta _{\widehat{\lambda }}\widehat{\omega }_{\mu }^{AB}=\partial _{\mu }%
\widehat{\lambda }^{AB}+\widehat{\omega }_{\mu }^{AC}*\widehat{\lambda }%
^{CB}-\widehat{\lambda }^{AC}*\widehat{\omega }_{\mu }^{CB}
\]
To solve this equation we first write
\begin{eqnarray*}
\widehat{\omega }_{\mu }^{AB} &=&\omega _{\mu }^{AB}+\omega _{\mu }^{\prime
AB}\left( \omega \right) \\
\widehat{\lambda }^{AB} &=&\lambda ^{AB}+\lambda ^{^{\prime }AB}\left(
\lambda ,\omega \right)
\end{eqnarray*}
where $\omega _{\mu }^{\prime AB}\left( \omega \right) $ and $\lambda
^{^{\prime }AB}\left( \lambda ,\omega \right) $ are functions of $\theta $,
and then substitute into the varriational equation to get \cite{SW}
\begin{eqnarray*}
&&\omega _{\mu }^{\prime AB}\left( \omega +\delta \omega \right) -\omega
_{\mu }^{\prime AB}\left( \omega \right) \\
&=&\partial _{\mu }\lambda ^{^{\prime }AB}+\omega _{\mu }^{AC}\lambda
^{\prime CB}-\lambda ^{\prime AC}\omega _{\mu }^{CB}+\omega _{\mu }^{\prime
AC}\lambda ^{CB}-\lambda ^{AC}\omega _{\mu }^{\prime CB} \\
&&+\frac{i}{2}\theta ^{\nu \rho }\left( \partial _{\nu }\omega _{\mu
}^{AB}\partial _{\rho }\lambda ^{CB}+\partial _{\rho }\lambda ^{AC}\partial
_{\nu }\omega _{\mu }^{CB}\right)
\end{eqnarray*}
This equation is solved, to first order in $\theta $, by
\begin{eqnarray*}
\widehat{\omega }_{\mu }^{AB} &=&\omega _{\mu }^{AB}-\frac{i}{4}\theta ^{\nu
\rho }\left\{ \omega _{\nu },\,\partial _{\rho }\omega _{\mu }+R_{\rho \mu
}\right\} ^{AB}+O(\theta ^{2}) \\
\widehat{\lambda }^{AB} &=&\lambda ^{AB}+\frac{i}{4}\theta ^{\nu \rho
}\left\{ \partial _{\nu }\lambda ,\,\omega _{\rho }\right\} ^{AB}+O(\theta
^{2})
\end{eqnarray*}
where we have defined the anticommutator $\left\{ \alpha ,\,\beta \right\}
^{AB}\equiv \alpha ^{AC}\beta ^{CB}+\beta ^{AC}\alpha ^{CB}.$ With this it
is possible to derive the differential equation that govern the dependence
of the deformed fields on $\theta $ to all orders
\[
\delta \widehat{\omega }_{\mu }^{AB}\left( \theta \right) =-\frac{i}{4}%
\theta ^{\nu \rho }\left\{ \widehat{\omega }_{\nu },_{*}\,\partial _{\rho }\,%
\widehat{\omega }_{\mu }+\widehat{R}_{\rho \mu }\right\} ^{AB}
\]
with the products in the anticommutator given by the star product, and where
\[
\widehat{R}_{\mu \nu }^{AB}=\partial _{\mu }\widehat{\omega }_{\nu
}^{AB}-\partial _{\nu }\widehat{\omega }_{\mu }^{AB}+\widehat{\omega }_{\mu
}^{AC}*\widehat{\omega }_{\nu }^{CB}-\widehat{\omega }_{\nu }^{AC}*\widehat{%
\omega }_{\mu }^{CB}
\]

We are mainly interested in determining $\widehat{\omega }_{\mu }^{AB}\left(
\theta \right) $ to second order in $\theta $. This is due to the fact that
the deformed gravitational action is required to be hermitian. The undefomed
fields being real, then implies that all odd powers of $\theta $ in the
action must vanish. The above equation could be solved iteratively, by
inserting the solution to first order in $\theta $ in the differntial
equation and integrating it. The second order corrections in $\theta $ to $%
\widehat{\omega }_{\mu }^{AB}$ are
\begin{eqnarray*}
&&\frac{1}{32}\theta ^{\nu \rho }\theta ^{\kappa \sigma }\left( \left\{
\omega _{\kappa },\,2\left\{ R_{\sigma \nu },\,R_{\mu \rho }\right\}
-\left\{ \omega _{\nu },\left( D_{\rho }R_{\sigma \mu }+\partial _{\rho
}R_{\sigma \mu }\right) \,\right\} -\partial _{\sigma }\left\{ \omega _{\nu
,}\,\left( \partial _{\rho }\omega _{\mu }+R_{\rho \mu }\right) \right\}
\right\} ^{AB}\right.  \\
&&\qquad \qquad \quad \left. +\left[ \partial _{\nu }\omega _{\kappa
},\,\partial _{\rho }\left( \partial _{\sigma }\omega _{\mu }+R_{\sigma \mu
}\right) \right] ^{AB}-\left\{ \left\{ \omega _{\nu },\,\left( \partial
_{\rho }\omega _{\kappa }+R_{\rho \kappa }\right) \right\} ,\,\left(
\partial _{\sigma }\omega _{\mu }+R_{\sigma \mu }\right) \right\}
^{AB}\right)
\end{eqnarray*}
One problem remains of how to determine the dependence of the vierbein $%
\widehat{e}_{\mu }^{a}$ on the undefomred field as it is not a gauge field.
To resolve this problem we adopt the strategy of considering the field $%
e_{\mu }^{a}$ as the gauge field of the translational generator of the
inhomegenious Lorentz group, obtained through the contraction of the group $%
SO(4,1)$ to $ISO(3,1).$ This is done as follows. First define the $SO(4,1)$
gauge field $\omega _{\mu }^{AB}$ with the field strength
\[
R_{\mu \nu }^{AB}=\partial _{\mu }\omega _{\nu }^{AB}-\partial _{\nu }\omega
_{\mu }^{AB}+\omega _{\mu }^{AC}\omega _{\nu }^{CB}-\omega _{\nu
}^{AC}\omega _{\mu }^{CB}
\]
and let $A=a,5.$ Define $\omega _{\mu }^{a5}=ke_{\mu }^{a}$. Then we have
\begin{eqnarray*}
R_{\mu \nu }^{ab} &=&\partial _{\mu }\omega _{\nu }^{ab}-\partial _{\nu
}\omega _{\mu }^{ab}+\omega _{\mu }^{ac}\omega _{\nu }^{cb}-\omega _{\nu
}^{ac}\omega ^{cb}+k^{2}\left( e_{\mu }^{a}e_{\nu }^{b}-e_{\nu }^{a}e_{\mu
}^{b}\right)  \\
R_{\mu \nu }^{a5} &\equiv &kT_{\mu \nu }^{a}=k\left( \partial _{\mu }e_{\nu
}^{a}-\partial _{\nu }e_{\mu }^{a}+\omega _{\mu }^{ac}e_{\nu }^{c}-\omega
_{\nu }^{ac}e_{\mu }^{c}\right)
\end{eqnarray*}
The contraction is done by taking the limit $k\rightarrow 0.$ By imposing
the condition $T_{\mu \nu }^{a}=0$ one can solve for $\omega _{\mu }^{ab}$
in terms of $e_{\mu }^{a}.$ We shall adopt a similar strategy in the
deformed case. We write $\widehat{\omega }_{\mu }^{a5}=k\widehat{e}_{\mu
}^{a}$ and $\widehat{\omega }_{\mu }^{55}=k\widehat{\phi }_{\mu }.$ We shall
only impose the condition $T_{\mu \nu }^{a}=0$ and not $\widehat{T}_{\mu \nu
}^{a}=0$ because we are not interested in $\phi _{\mu }$ which will drop out
in the limit $k\rightarrow 0.$ The result for $\widehat{e}_{\mu }^{a}$ in
the limit $k\rightarrow 0$ is
\begin{eqnarray*}
\widehat{e}_{\mu }^{a} &=&e_{\mu }^{a}-\frac{i}{4}\theta ^{\nu \rho }\left(
\omega _{\nu }^{ac}\partial _{\rho }e_{\mu }^{c}+\left( \partial _{\rho
}\omega _{\mu }^{ac}+R_{\rho \mu }^{ac}\right) e_{\nu }^{c}\right)  \\
&&+\frac{1}{32}\theta ^{\nu \rho }\theta ^{\kappa \sigma }\left( 2\left\{
R_{\sigma \nu },R_{\mu \rho }\right\} ^{ac}e_{\kappa }^{c}-\omega _{\kappa
}^{ac}\left( D_{\rho }R_{\sigma \mu }^{cd}+\partial _{\rho }R_{\sigma \mu
}^{cd}\right) e_{\nu }^{d}\right.  \\
&&-\left\{ \omega _{\nu },\left( D_{\rho }R_{\sigma \mu }+\partial _{\rho
}R_{\sigma \mu }\right) \right\} ^{ad}e_{\kappa }^{d}-\partial _{\sigma
}\left\{ \omega _{\nu },\,\left( \partial _{\rho }\omega _{\mu }+R_{\rho \mu
}\right) \right\} ^{ac}e_{\kappa }^{c} \\
&&-\omega _{\kappa }^{ac}\partial _{\sigma }\left( \omega _{\nu
}^{cd}\partial _{\rho }e_{\mu }^{d}+\left( \partial _{\rho }\omega _{\mu
}^{cd}+R_{\rho \mu }^{cd}\right) e_{\nu }^{d}\right) +\partial _{\nu }\omega
_{\kappa }^{ac}\partial _{\rho }\partial _{\sigma }e_{\mu }^{c} \\
&&-\partial _{\rho }\left( \partial _{\sigma }\omega _{\mu }^{ac}+R_{\sigma
\mu }^{ac}\right) \partial _{\nu }e_{\kappa }^{c}-\left\{ \omega _{\nu
},\,\left( \partial _{\rho }\omega _{\kappa }+R_{\rho \kappa }\right)
\right\} ^{ac}\partial _{\sigma }e_{\mu }^{c} \\
&&-\left. \left( \partial _{\sigma }\omega _{\mu }^{ac}+R_{\sigma \mu
}^{ac}\right) \left( \omega _{\nu }^{cd}\partial _{\rho }e_{\kappa
}^{d}+\left( \partial _{\rho }\omega _{\kappa }^{cd}+R_{\rho \kappa
}^{cd}\right) e_{\nu }^{d}\right) \right) +O\left( \theta ^{3}\right)
\end{eqnarray*}

At this point, it is possible to determine the deformed curvature and use it
to calculate the deformed action given by
\[
\int d^{4}x\sqrt{\widehat{e}}*\,\widehat{e}_{*a}^{\mu }*\widehat{R}_{\mu \nu
}^{ab}*\left( e_{*b}^{\nu }\right) ^{\dagger }*\sqrt{\widehat{e}}^{\dagger }
\]
Notice that this action is hermitian including the measure. We have defined $%
\widehat{e}=\det \left( \widehat{e}_{\mu }^{a}\right) ,$ and the inverse
vierbein by
\[
\widehat{e}_{*a}^{\mu }*\widehat{e}_{\mu }^{b}=\delta _{a}^{b}
\]
This will determine the inverse deformed vierbein as an expansion in $\theta
$. By writing
\begin{eqnarray*}
\widehat{e}_{\mu }^{a} &=&e_{\mu }^{a}+i\theta ^{\nu \rho }e_{\mu \nu \rho
}^{a}+\theta ^{\nu \rho }\theta ^{\kappa \sigma }e_{\mu \nu \rho \kappa
\sigma }^{a}+O(\theta ^{3}) \\
\widehat{e}_{*a}^{\mu } &=&e_{a}^{\mu }+i\theta ^{\nu \rho }e_{a\nu \rho
}^{\mu }+\theta ^{\nu \rho }\theta ^{\kappa \sigma }e_{a\nu \rho \kappa
\sigma }^{\mu }+O(\theta ^{3})
\end{eqnarray*}
where $e_{\mu \nu \rho }^{a}$ and $e_{\mu \nu \rho \kappa \sigma }^{a}$ can
be read from the expansion of $\widehat{e}_{\mu }^{a}$ to second order in $%
\theta $, one finds that
\begin{eqnarray*}
e_{a\nu \rho }^{\mu } &=&-e_{b}^{\mu }\left( e_{a}^{\kappa }e_{\kappa \nu
\rho }^{b}+\frac{1}{2}\partial _{\nu }e_{a}^{\kappa }\partial _{\rho
}e_{\kappa }^{b}\right)  \\
e_{a\nu \rho \kappa \sigma }^{\mu } &=&-e_{b}^{\mu }\left( e_{a\nu \rho
}^{\alpha }e_{\alpha \kappa \sigma }^{b}+e_{a}^{\alpha }e_{\alpha \nu \rho
\kappa \sigma }^{b}-\frac{1}{4}\partial _{\nu }\partial _{\kappa
}e_{a}^{\alpha }\partial _{\rho }\partial _{\sigma }e_{\alpha }^{b}-\frac{1}{%
2}\partial _{\nu }e_{a}^{\alpha }\partial _{\rho }e_{\alpha \kappa \sigma
}^{b}+\frac{1}{2}\partial _{\rho }e_{\alpha }^{b}\partial _{\nu }e_{a\kappa
\sigma }^{\alpha }\right)
\end{eqnarray*}

It is legitimate to question the meaning of the star product under general
coordinate transformations, and whether this action is diffeomorphism
invariant. Afterall, the original definition of the star product assumes
that the commutator
\[
\left[ x^{\mu },\,x^{\nu }\right] =i\theta ^{\mu \nu }
\]
is constant. However, under diffeomorphism transformations, $\theta ^{\mu
\nu }$ becomes a function of $x$, and one has to generalize the definition
of the star product to be applicable for a general manifold. The
prescription for doing this was given by Kontsevich \cite{kont}. The star
product is then defined by
\begin{eqnarray*}
f*g &=&fg+
\rlap{\protect\rule[1.1ex]{.325em}{.1ex}}h%
B_{1}\left( f,g\right) +
\rlap{\protect\rule[1.1ex]{.325em}{.1ex}}h%
^{2}B_{2}\left( f,g\right) +\cdots  \\
&=&fg+
\rlap{\protect\rule[1.1ex]{.325em}{.1ex}}h%
\alpha ^{ab}\partial _{a}f\partial _{b}g+\frac{1}{2}
\rlap{\protect\rule[1.1ex]{.325em}{.1ex}}h%
^{2}\alpha ^{ab}\alpha ^{cd}\partial _{a}\partial _{c}f\partial _{b}\partial
_{d}g+ \\
&&+\frac{1}{3}
\rlap{\protect\rule[1.1ex]{.325em}{.1ex}}h%
^{2}\alpha ^{as}\partial _{s}\alpha ^{bc}\left( \partial _{a}\partial
_{b}f_{c}\partial g+\partial _{a}\partial _{b}g\partial _{c}f\right) +O(
\rlap{\protect\rule[1.1ex]{.325em}{.1ex}}h%
^{3})
\end{eqnarray*}
where
\[
\alpha ^{ab}=\theta ^{\mu \nu }\partial _{\mu }z^{a}\partial _{\nu
}z^{b}+O(\theta ^{3})
\]

Therefore under diffeomorphisms the star product transforms according to
\begin{eqnarray*}
\ast &\rightarrow &*^{^{\prime }} \\
f^{\prime }(
\rlap{\protect\rule[1.1ex]{.325em}{.1ex}}h%
) &=&Df(
\rlap{\protect\rule[1.1ex]{.325em}{.1ex}}h%
) \\
f^{\prime }*^{\prime }g^{\prime } &=&D(D^{-1}f^{\prime }*D^{-1}g^{\prime })
\\
D &=&1+\sum\limits_{i=1}
\rlap{\protect\rule[1.1ex]{.325em}{.1ex}}h%
D_{i}
\end{eqnarray*}
In this case \cite{zotov}
\[
D=1-\frac{
\rlap{\protect\rule[1.1ex]{.325em}{.1ex}}h%
^{2}}{4}\theta ^{\mu \nu }\theta ^{\rho \sigma }\left( \partial _{\mu
}\partial _{\rho }z^{a}\partial _{\nu }\partial _{\sigma }z^{b}\partial
_{a}\partial _{b}+\frac{2}{3}\left( \partial _{\mu }\partial _{\rho
}z^{a}\partial _{\nu }z^{b}\partial _{\sigma }z^{c}\right) \partial
_{a}\partial _{b}\partial _{c}\right) +O(
\rlap{\protect\rule[1.1ex]{.325em}{.1ex}}h%
^{4})
\]

It is thus possible to define the star product to be invariant under
diffeomorphism transformations..

\section{The action to second order in $\theta $}

To determine the action to second order in $\theta $, we first write
\[
\widehat{R}_{\mu \nu }^{ab}=R_{\mu \nu }^{ab}+i\theta ^{\rho \tau }R_{\mu
\nu \rho \tau }^{ab}+\theta ^{\rho \tau }\theta ^{\kappa \sigma }R_{\mu \nu
\rho \tau \kappa \sigma }^{ab}+O(\theta ^{3})
\]
where
\begin{eqnarray*}
R_{\mu \nu \rho \tau }^{ab} &=&\partial _{\mu }\omega _{\nu \rho \tau
}^{ab}+\omega _{\mu }^{ac}\omega _{\nu \rho \tau }^{ac}+\omega _{\mu \rho
\tau }^{ac}\omega _{\nu }^{cb}-\frac{1}{2}\partial _{\rho }\omega _{\mu
}^{ac}\partial _{\tau }\omega _{\upsilon }^{cb}-\mu \leftrightarrow \nu  \\
R_{\mu \nu \rho \tau \kappa \sigma }^{ab} &=&\partial _{\mu }\omega _{\nu
\rho \tau \kappa \sigma }^{ab}+\omega _{\mu }^{ac}\omega _{\nu \rho \tau
\kappa \sigma }^{ac}+\omega _{\mu \rho \tau \kappa \sigma }^{ac}\omega _{\nu
}^{cb}-\omega _{\mu \rho \tau }^{ac}\omega _{\nu \kappa \sigma }^{ac} \\
&&-\frac{1}{4}\partial _{\rho }\partial _{\kappa }\omega _{\mu
}^{ac}\partial _{\tau }\partial _{\sigma }\omega _{\upsilon }^{cb}-\mu
\leftrightarrow \nu
\end{eqnarray*}
With this we can now expand
\begin{eqnarray*}
\,\widehat{e}_{*a}^{\mu }*\widehat{R}_{\mu \nu }^{ab}*\left( e_{*b}^{\nu
}\right) ^{\dagger } &=&R \\
&&+\theta ^{\rho \tau }\theta ^{\kappa \sigma }\left( e_{a}^{\mu }R_{\mu \nu
\rho \tau \kappa \sigma }^{ab}e_{b}^{\nu }+e_{a\rho \tau \kappa \sigma
}^{\mu }R_{\mu \nu }^{ab}e_{b}^{\nu }+e_{a}^{\mu }R_{\mu \nu }^{ab}e_{b\rho
\tau \kappa \sigma }^{\nu }\right.  \\
&&\qquad \qquad \left. -e_{a\rho \tau }^{\mu }R_{\mu \nu }^{ab}e_{b\kappa
\sigma }^{\nu }-e_{a\rho \tau }^{\mu }R_{\mu \nu \kappa \sigma
}^{ab}e_{b}^{\nu }-e_{a}^{\mu }R_{\mu \nu \rho \tau }^{ab}e_{b\kappa \sigma
}^{\nu }\right) +O(\theta ^{4})
\end{eqnarray*}
Notice that the odd powers of $\theta $ cancel because of the hermiticity of
the above expression and reality of the undeformed fields. The expansion of
the determinant is
\[
\det \left( \widehat{e}_{\mu }^{a}\right) =1+i\theta ^{\rho \tau }e_{a}^{\mu
}e_{\mu \rho \tau }^{a}+\frac{1}{2}\theta ^{\rho \tau }\theta ^{\kappa
\sigma }\left( e_{\mu \rho \tau }^{a}e_{\nu \kappa \sigma }^{a}e_{c}^{\mu
}e_{c}^{\nu }-e_{\mu \rho \tau }^{a}e_{\nu \kappa \sigma }^{b}e_{a}^{\mu
}e_{b}^{\nu }\right) +O(\theta ^{3})
\]
This completes the action to second order in $\theta $. Of course the actual
expression obtained after substiuting for the fields $e_{a\rho \tau }^{\mu },
$ $e_{a\rho \tau \kappa \sigma }^{\mu }$, $\omega _{\mu \rho \tau }^{ab}$
and $\omega _{\mu \rho \tau \kappa \sigma }^{ab}$ is very complicated, and
it is not clear whether one can associate a geometric structure with it. One
can, however, take this expression and study the deformations to the
graviton propagator, which will receive $\theta ^{2}$ corrections. It is an
interesting question to study the effect of the deformation on the
renormalizability of the theory. This project will have to be handled by
using a computer program for algebraic manipulations.

\section{Conclusion}

In this work we have shown that it is possible to deform Einstein's gravity
without introducing new fields. The idea is based on the gauging of the
noncommutative gauge group $ISO(3,1)$ and using the Seiberg-Witten map to
express the deformed fields in terms of the undeformed ones. The reality of
the undeformed fields and the hermiticity of the action implies that the
lowest order correction to Einstein's action is second order in the
deformation parameter. This makes the form of the corrections fairly
complicated. It is, however, possible to use perturbation theory to
determine the modified graviton propagator as well as the vertex operators,
to second order in $\theta $. It is an interesting problem to study the
renormalizability of the theory and the effects of the deformation parameter
on the infrared and ultraviolet divergencies. Performing these calculations
will be left for future work.

\section{Acknowledgments}

I would like to thank the Alexander von Humboldt Foundation for support
through a research award. I would also like to thank Slava Mukhanov for
hospitality at the Ludwig-Maxmiilans University in M\"{u}nich where part of
this work was done.

\end{document}